\documentclass[apj]{emulateapj}
\shorttitle{the PACMan Review Tool}
\shortauthors{Strolger et al.}

\usepackage{wasysym,ulem,color}
\usepackage{natbib}
\begin{document}

\title{The Proposal Auto-Categorizer and Manager\\ for Time Allocation Review at Space Telescope Science Institute}
\author{Louis-Gregory~Strolger\altaffilmark{1}, Sophia~Porter\altaffilmark{1,2}, Jill~Lagerstrom\altaffilmark{1}, Sarah~Weissman\altaffilmark{1}, I.~Neill~Reid\altaffilmark{1}, and Michael~Garcia\altaffilmark{3}}
\altaffiltext{1}{Space Telescope Science Institute, Baltimore, MD, 21218}
\altaffiltext{2}{Johns Hopkins University, Baltimore, MD 21218}
\altaffiltext{3}{NASA Headquarters, Astrophysics Division, Washington, DC, 20546}

\begin{abstract}
 The Proposal Auto-Categorizer and Manager (PACMan) tool was written to respond to concerns on subjective flaws and potential biases in some aspects of the proposal review process for time allocation for the {\it Hubble Space Telescope} (HST), and to partially alleviate some of the anticipated additional workload  from the {\it James Webb Space Telescope} (JWST) proposal review. PACMan is essentially a mixed-method Naive Bayesian spam filtering routine, with multiple pools representing scientific categories, that utilizes the Robinson method for combining token (or word) probabilities. PACMan was trained to make similar programmatic decisions in science category sorting, panelist selection, and proposal-to-panelists assignments to those made by individuals and committees in the Science Policies Group (SPG) at Space Telescope Science Institute. Based on training from the previous cycle's proposals, PACMan made the same science category assignments for proposals in Cycle 24 as did the SPG, an average of 87\% of the time. Tests for similar science categorizations, based on training using proposals from additional cycles, show that this accuracy can be further improved, to the $>95\%$ level. This tool will be used to augment or replace key functions in the TAC review processes in future HST and JWST cycles.
\end{abstract}	
	
\section{Introduction}
The purpose of the peer review process in the allocation of telescope time is to select proposals of the highest scientific merit. Similar general processes are used worldwide, for both ground and space-based telescopes. For the {\it Hubble Space Telescope} (HST) review, the Science Policies Group (SPG) at Space Telescope Science Institute (STScI) arranges the annual peer review, enlisting over 150 international members of the astronomy community in a six-month review process on approximately 1,100 proposals that culminates in ranked recommendations to the STScI Director at a face-to-face Time Allocation Committee (TAC) meeting. However, as the launch of the {\it James Webb Space Telescope} (JWST) nears, the process which already sees tremendous proposal pressure is on the verge of essentially doubling the number of proposals it receives and must accommodate. Arranging a fair and thorough review of all submissions now appears to be a daunting and ever more time-consuming endeavor. There is clearly a need to develop tools and strategies to streamline the review processes, and to do so without increasing the potential for explicit and implicit biases typically encountered in pressured reviews. 

The Proposal Auto-Categorizer and Manager (PACMan) is a Naive Bayesian classifier developed to assist the SPG in its principal task:~recruiting qualified reviewers for each proposal. As will be discussed further below, the tool automatically performs the same key tasks as the SPG in preparing for the review: it sorts proposals into pre-defined scientific categories and identifies people to serve on panels based on those same science areas, based on the content within the proposal and publication record of reviewers in those scientific areas. PACMan will be implemented in upcoming HST and JWST cycles to support parts of process for constructing the time allocation reviews. Here we provide an overview of PACMan and present the results of its training and testing on up to five cycles of proposals with HST. 

The PACMan package, including algorithm, the Cycle 23-trained library, and ancillary code, is available on GitHub~\citep{SP2017} under the terms of the MIT license. The training data is, however, proprietary and is not included in the package.

With work, we believe PACMan could be used to directly match specific proposals to a pool of expert reviewers, potentially bypassing assembly by science category, which could revolutionize the way in which we do time allocation reviews at STScI, and potentially elsewhere.

This paper is organized as follows. In Section 2 we provide an overview of the current proposal review process, and discuss some areas of subjective biases which are of concern. In Section 3 we describe PACMan's mixed-method, Naive Bayesian approach, and how it is used to train and test corpora for categorization. in Section 4 we show the results of PACMan testing on Cycle 24 proposals and panelists, as well as results to tests over five cycles of proposal data, from Cycle 19 to 23. Finally, in Section 5 we discuss future directions for development of the tool, for use at STScI and elsewhere, in proposal review assignments.

\section{The Proposal Review Process, Subjective Flaws, and Potential Biases}
The time allocation process for HST~is similar in several respects to other processes for the allocation of resources in astronomy, but with some important differences that are worth mentioning, especially in the consideration of automating this part of the process. Logistical issues in arranging the TAC meeting, such as travel arrangements for international reviewers, necessitate composing the review panels several months in advance of the cycle proposal deadline, which of course means reviewers are selected well before proposals are received. While this  seems somewhat counter-intuitive, the process works in large part because of well-defined science categories--- themes which encapsulate related areas of expertise and facilitate the sorting of proposals for review. These categories, and their related scientific keywords, are defined with other proposing instructions in the release of the Call for Proposals.

\subsection{Proposal Sorting}
The SPG recruits reviewers (or panelists) based on their expertise in each of these science  categories, and proposers, at the time of submission, indicate which of the categories best describes their science at a fundamental level. As an aid, categories are supplemented by scientific keywords that provide further specificity of proposal topics and better define panelists' areas of expertise, and at some level, the matching of proposals to panelists.  And so, while this proposal sorting process may not necessarily be optimized for matching individual reviewers to specific proposals, as is done in the peer review of articles to science journals, this expertise matching from the use of categories and keywords assures the recruitment of subject-appropriate reviewers, well-suited for a broad range of likely to be proposed subject areas.

Proposal sorting, as described above, would ideally require little effort to implement, as most of the work in matching is done by the proposers and the selected panelists. However, experience has shown that category names and keywords can often be misunderstood, or misused, or simply inadequate for describing the science of a given proposal, and require SPG intervention to sort properly. 

\begin{figure*}[t] 
   \centering
   \includegraphics[width=5.5in]{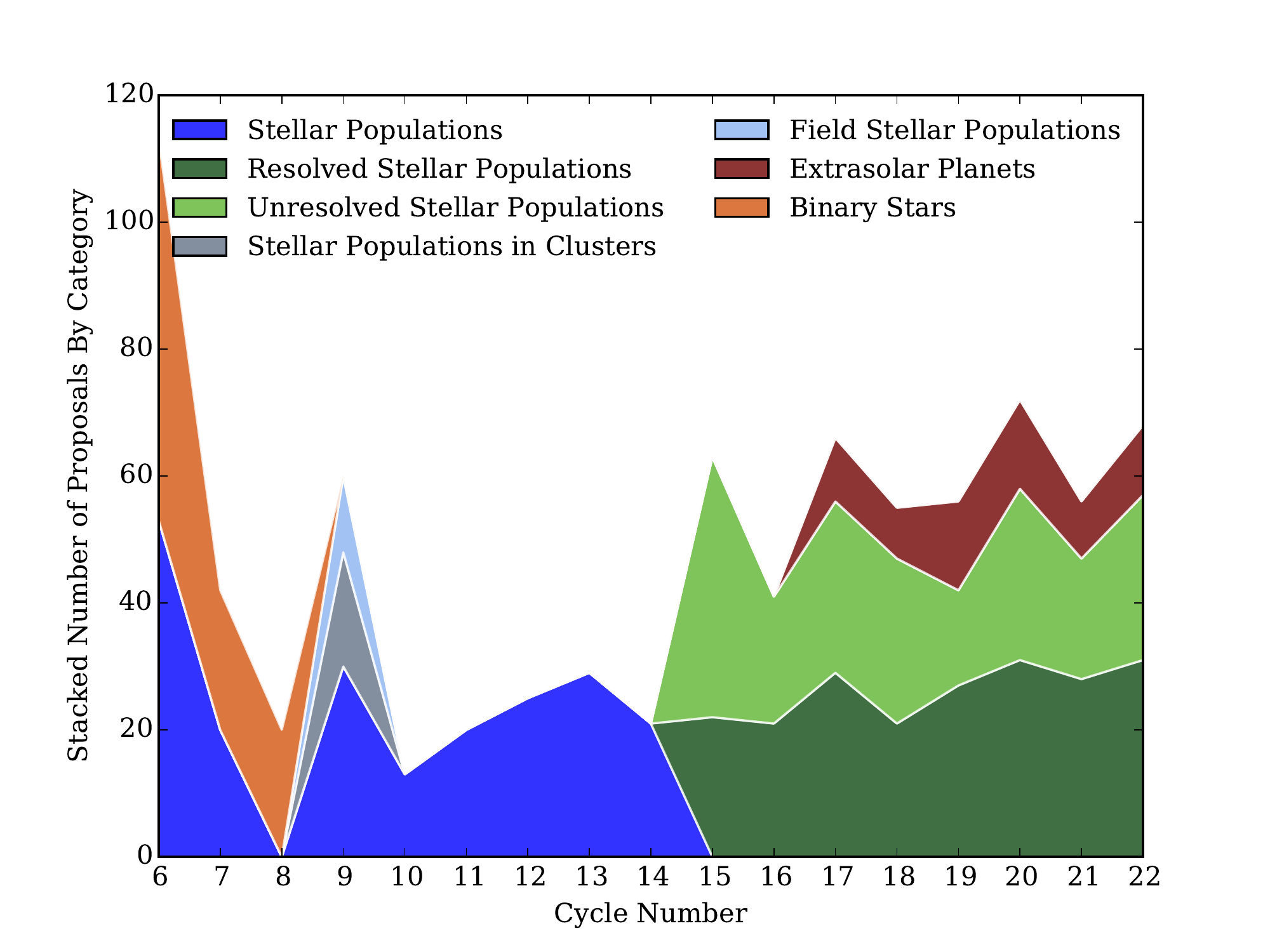}
   \caption{\footnotesize The frequency, normalized, with which a few science category names were used in the HST TAC process, from Cycle 6 (1995) to Cycle 22 (2014).}
   \label{fig:oldcats}
\end{figure*}

An important aspect of the ambiguity is exemplified by Figure~\ref{fig:oldcats}, which shows the names and use of a few example proposal science categories over a twenty year period. As is well exemplified by the {\it Stellar Populations} category, topics that are too broad become inundated, and are pressured to split into separate, more specific categories to balance the load on reviewers in said categories. These splits are not always obvious, and there have been missteps---  delineating {\it Stellar Populations in Clusters} from those in the field in Cycle 8 was immediately undone in the proceeding Cycle 9. But eventually this load-balancing pressure finds an appropriate split--- in this case to the {\it Resolved} and {\it Unresolved Stellar Populations} categories that have remained in use since Cycle 15. 

Categories also emerge or fade in response to emergent or less active fields, or with new or diminished capabilities with HST. For example, {\it Extra Solar Planets} is new as of Cycle 17, the first cycle after the HST-SM4 servicing mission installed Wide Field Camera 3 (WFC3), with near-infrared spectroscopic capabilities useful for probing exoplanet atmospheres. Similarly, the Faint Object Spectrograph (FOS) and the Goddard High Resolution Spectrograph (GHRS) were both quite popular in the spectroscopy of binary stars, but were replaced in HST-SM2 by the Near Infrared Camera and Multi-Object Spectrometer (NICMOS) and the Space Telescope Imaging Spectrograph (STIS) just prior to Cycle 7. Post Cycle 7, proposals in topics previously covered in {\it Binary Stars} have been reduced enough to be subsumed in more general stars and stellar physics categories. 

While the intention behind the changes is to focus or otherwise improve the proposal review process, it is well appreciated that these and other frequent changes may have created confusion, making it less obvious to which category a proposer should  submit, or where panelist's areas of expertise lie. 

There has been a similar issue with scientific keywords associated with proposal assignments, as they, too, change with category changes. They also have not been unique terms specific to a given category. Despite explanation in documentation on the intended meaning of keywords, and descriptions of proposal sorting criteria, roughly $10\%$ of proposals are not well suited to the categories their proposers selected. Their sorting into proper categories is a process performed manually by the SPG, which is, not surprisingly, a time-consuming endeavor.

\subsection{Panel Recruitment}
As the SPG recruits $>150$ scientists for the review each year, it does so with just a few restrictions and recruiting goals in mind. We attempt to recruit astronomers in at least early-to-mid career stages, and as a minimum require panelists to have a Ph.D. in astronomy or closely related fields, with at least three years of professional experience. Some familiarity with HST and its instrumentation is desired, but not required, as panelists will be principally taxed with evaluating scientific merit, not technical feasibility. \cite{Reid2014} addressed the challenges in reducing conscious and unconscious bias in these reviews, and those ends, we also strive to at least achieve gender balance on the panels, with serious consideration to the geographical, racial, and ethnic diversity of the panelists.

  Yet, despite these few restrictions, service on the TAC and panels has strongly weighted towards astronomers who have previously served, on this committee or others for STScI, or have received observing time in recent cycles. The SPG record-keeping has greatly improved in recent years, facilitated by the ProPer~\footnote{The Proposal/Person (ProPer) database, {\tt https://proper.stsci.edu/proper/}.}  database, allowing for a better tracking of who has served on STScI committees, in what capacities and when, increasing our awareness of latent recruiting biases. The SPG aims to keep repeated service on the TAC infrequent, while keeping in mind that some institutional memory can be advantageous. Nonetheless, recruitment for TAC panel service has been somewhat limited, as the pools we query remain generally insular.

Ultimately, PACMan is intended to reduce subjectivity in proposal classification, and biases in panelist selection, while reducing SPG effort. We defined success for a desired automated routine as one which makes decisions in proposal sorting and review panel assembly to those of the SPG at least 90\% of the time. It is also possible that observing these objective trends in assignments over time might motivate the creation of more permanent categories that better and more precisely reflect modern science.

\section{How the Bayesian Method Works}
The problem of classifying a corpus based on the words it contains is a rather common one, with numerous solutions in machine learning. Among the simplest is the ``bag of words'' approach-- a Naive Bayesian classification tool that is simple yet effective in sorting text, and more complex types of data, into discrete categories. The technique is used in a broad range of sorting tasks, from facial recognition~\citep{Moghaddam2000}, to the interpretation of medical data~\citep{AlAidaroos2012}. But Naive Bayes' are seen most often in junk-mail filtering. The basic actions of a junk-mail, or spam, filter are to (a) determine the probability that each word in a corpus is a ``spam-word'' based on the frequency of its appearance in corpora already identified as spam, and then (b) determine the probability of a new corpus being spam based on a combination of the probabilities assigned to the spam-words used in the corpus. These same actions can be generalized to a wider set of classifications (or pools), not just spam or not-spam, by taking into consideration a wider set of combined probabilities trained from a larger selection of known corpora.

In a more specific application of Bayes' method, given a token (or word), $w_i$, the likelihood, $P_{ij}$, that the corpus it's drawn from belongs in a specific pool (or scientific category), $p_j$, is given by:

\begin{equation}
	P_{ij}\equiv P(p_j|w_i) = \frac{P(w_i|p_j)\cdot P(p_j)}{\sum_i^N [P(w_i|p_j)\cdot P(p_j)]}.
	\label{eqn:pij}
\end{equation}

\noindent Here, we assume the probability of a word belonging to a specified category, $P(w_i|p_j)$, or {\it trigger word}, is simply the frequency of use of the word in previously identified training corpora. There are no words that are assigned weight disproportionate to their frequency of use in a particular category. The value is weighed by the sum of probabilities for all words in the training set for the category, as shown by the denominator in Equation~\ref{eqn:pij}.  The term $P(p_j)$, which represents any priors preferentially favoring one category over any other, is set to $P(p_j)=1/N_{\rm pools}$ in a non-biased approach, as we assume all categories are equally likely. 

The overall probability that an entire corpus belongs in a given category-pool, $P_j$, depends on how the probabilities for each word-token are combined. If one assumes the words in a corpus are independent events, i.e., tokens like ``proto'' do not have a predilection for other tokens like ``stars'', then the word probabilities from Equation~\ref{eqn:pij} could be combined in a straight-forward, hence naive, product-combination derived directly from Bayes' theorem. For example, using Paul Graham's method~\citep{zdz2005}, 

\begin{equation}
	P_j=\frac{\prod_i^N P_{ij}}{\prod_i^N P_{ij}+\prod_i^N (1-P_{ij})}.
	\label{eqn:pj}
\end{equation}

\noindent However, implementing this approach leads to rather draconian results, with probabilities either very close to zero, or close to unity, with little to no room for evaluating relative likelihoods. One could interpret this as falsely (or naively) assuming token independence-- not accurately reflecting the interrelationships of individual word-tokens.

\begin{figure}[t] 
   \centering
   \includegraphics[width=3.5in]{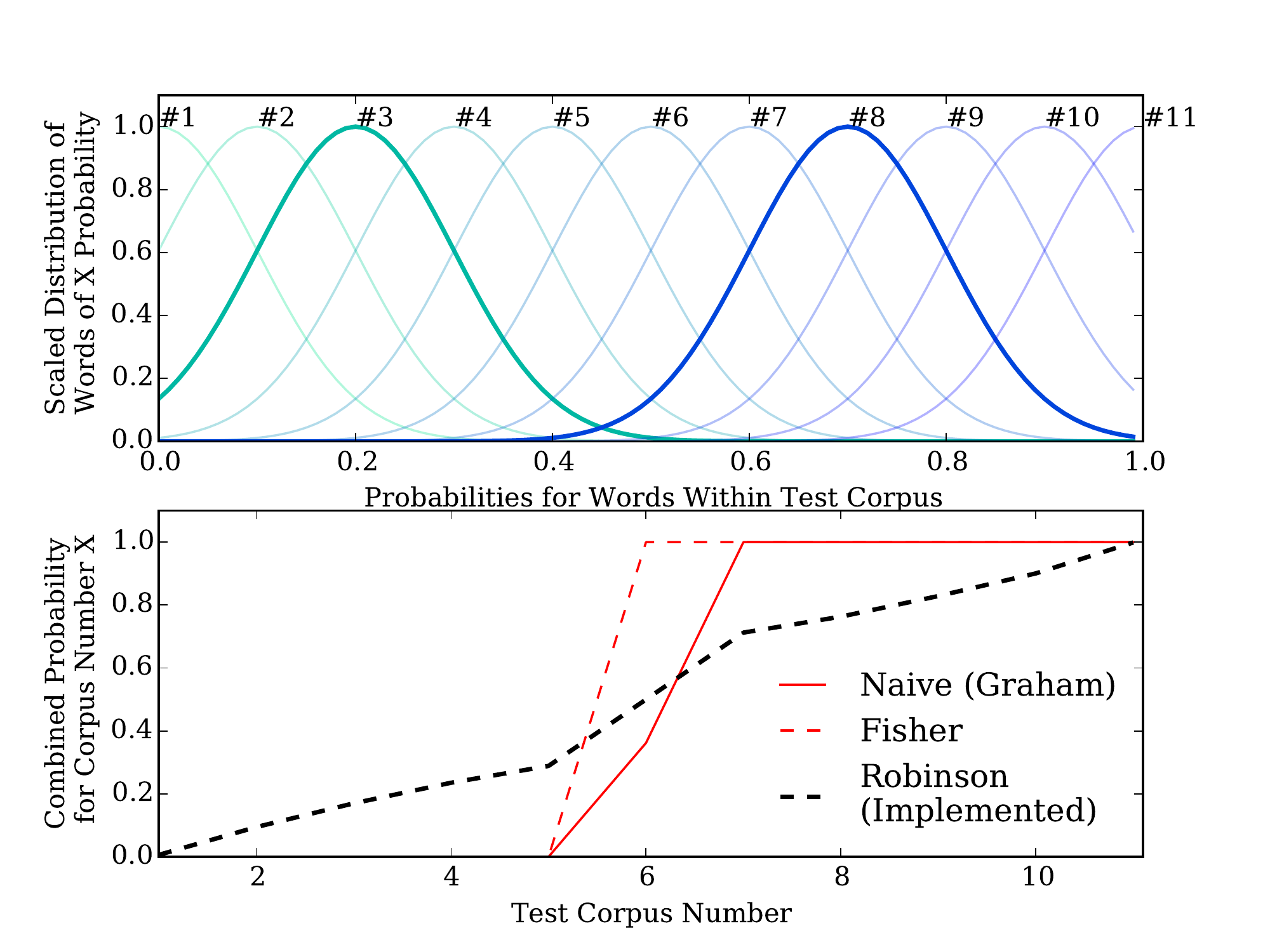}
   \caption{\footnotesize An illustration of different combined probability results. {\it Upper panel:} The distributions of word probabilities from 11 mock corpora, tested against a mock category, each exhibiting a small range of appropriateness to the trained category. {\it Lower panel:} The comparison of token probability combination methods on the mock corpora.}
   \label{fig:p_combo}
\end{figure}

This effect is illustrated in a example shown in Figure~\ref{fig:p_combo}. Here, we test 11 mock corpora against training for a given mock category. The distributions of the word probabilities, $P(w_i|p_j)$ (shown in the upper panel), are all made Gaussian for simplicity, and are set such that some are composed entirely of poor word probabilities for the category, at the left of the figure, and others are essentially just trigger words, to the right. The bottom panel shows what would result for each mock corpus if the naive combination of probabilities from Equation~\ref{eqn:pj} was used---  either very high combined probability (unity) or very low (zero), with a very narrow transition region, as shown by the solid red line. Similar methods such as the Fisher's Inverse Chi-Square test~\citep{zdz2005} have similar results, which are shown by dashed red line in Figure~\ref{fig:p_combo}. It is, perhaps, not surprising the utility of both methods to their primary application in distinguishing between two disparate pools--- separating junk e-mail from important correspondence.

There should, however, be some gray area in categorizing proposals and reviewers as science can (and often does) have bearing to more than one subject area, and expertise can be quite broad.  A method yielding more granular results is Robinson's Geometric Mean test~\citep{zdz2005}, which determines the level of belonging to a category by 

\begin{equation}
	P_j' = 1 - \prod_i^N (1-P_{ij})^{(1/N)}, 
\end{equation}
\noindent and combines this with the probability of not belonging to the category,

\begin{equation}
	Q_j = 1 - \prod_i^N P_{ij}^{(1/N)},
\end{equation}
\noindent using 

\begin{equation}
	P_j = \frac{1}{2}\biggl[1+\frac{(P_j'-Q_j)}{(P_j'+Q_j)}\biggr].
	\label{eqn:pj1}
\end{equation}

\noindent This method yields more gradual results, nearly correlated with the mean corpus word probability, as shown by the black line at the bottom of Figure~\ref{fig:p_combo}. But, by it's nature, this is not as effective at reducing noise as the Graham method.
    
As discussed in~\cite{zdz2005}, the above mixed-methodology could be further improved in several known ways, including improving the tokenization by manually accounting for ``rare words'', or influential words that are used infrequently, but are also strong indicators for a specific characterization pool. Ignoring the impact of these words, as PACMan does, incorrectly weights test corpora toward more indifferent categorizations.

\subsection{Method for Training and Testing the Classification of Proposals}
Our primary goal has been to train PACMan on the HST proposals from Cycle 23, and apply the training to proposals from Cycle 24 to compare to the normal manual evaluation. As required by format, proposals contain a lot of information, much of which is not necessarily useful in determining what science area they pertain to. We assume that Abstracts and Scientific Justification sections are the most representative sections of the text for classification. The technical descriptions of observations do little to distinguish one science area from another, and most other proposal information is largely irrelevant, and therefore discarded. 

We use a few pre-processing strategies on corpora to disambiguate and improve the results of the training and testing afterwards. All punctuation, numeric characters, and special characters (e.g., the `{\AA}' symbol) are stripped from the text. We also remove words commonly used in any text (conjunctions, most verbs, pronouns, definitive articles, etc.) that have no distinguishing power in categorization, rather create a noise floor that must be overcome to start a categorization. Additional ``stop words'', or words that are commonly used in proposing of this type, e.g., `conclusive', `trend', and `robust', are similarly removed to reduce confusion.  Words are then reduced to a word root using a stemming algorithm~\citep{Porter1980} to remove non-distinguishing suffixes. For example, `galaxy' and `galaxies' are treated as the same root word, `galaxi', and are tallied and evaluated commonly.

We use the Divmod Reverend simple Bayesian classifier python package~\footnote {Reverend 0.4, {\tt https://pypi.python.org/pypi/Reverend}} to train corpora, building a dictionary of words and their frequency of use within a set with known categorizations, then test the dictionary on uncategorized corpora, utilizing the Robertson method. Categorization probabilities are treated as a raw score that are then normalized by the sum of raw scores over all categories, to require that each proposal fall at least one of the categories. A proposal should not  have normalized score of zero in all categories. The normalized scores for each corpus is then sorted, where the top score represents the most likely categorization. Categorizations in the top 60\%, generally only the first and second sorted category, but sometimes 3, are deemed to be reasonable `guesses' to the category. Those in the bottom 40\% are deemed to be poor categorizations for the proposal.

 \subsection{Method for Identifying and Categorizing Reviewers}\label{sec:reviewers}
Much of the initial framework for the PACMan algorithm was developed for tracking the science areas in which astronomers work, to provide demographic information to Infrastructure Study Groups, for the State of the Profession sections of the 2010 Astronomy Decadal Survey~\citep{2010Dec}.

The names of TAC panelists are queried in the SAO/NASA Astrophysics Data System~\footnote{http://adsabs.harvard.edu.} (ADS) to generate an `abstract bibliography' of the last ten years of refereed publications, which serves as the corpus for said author. The corpus is pre-processed using the same strategies described in the proposal process, and tested using the trained PACMan dictionary made in the proposal testing routine.  

Unlike what is done for proposals, we do not normalize the raw scores for panelists (authors), rather allow an interpretation of the raw score to determine how appropriate a panelist is in the tested area. This allows for expertise in several areas, without requiring specialization in a specific category. Additionally, we do not weight categorization scores by $h$-index, or by position in an author list, or number of authors, or any other similar metric which, preliminary tests have shown, could better distinguish and identify an individual's areas of expertise. Our justification for not further developing these weighted metrics is that abstract bibliographies should have about the same number of word tokens as proposals do to be evaluated to roughly same level of certainty. Most proposals have about 11,000 word-tokens in them. ADS abstract have a median of about 750 word-tokens, which multiplied by a median of 11 abstract per person, yields just over 8,000 word-tokens, which is moderately comparable to the size of proposals. Any further refining of the abstracts would lead to incomparable samples, on average. There is also, likely, a minimum number of word tokens required to provide a reliable classification. This lower limit would be highly dependent on what type of words were used, e.g., if they contain trigger words, and how many. For now this floor remains an unexplored caveat to this type of analysis.

The ADS service makes abstracts for most astronomy-related journals readily available, but unfortunately, there is no comparable service for obtaining more of a published text. As such, our tools are limited to just the words contained within a typical abstract. Also realizing that people become interested in multiple topics over time, we do attempt to keep some pertinency by restricting searches to only the last ten years of publication. But these two constraints necessarily limit any additional constraints or weights we can place on the data without risking incomparable corpus sizes, and thus reducing the certainty of the classifications.

\section{Results}
We have used PACMan to test the sorting of proposals, and the suitability of panelists to the received proposals, for the HST Cycle 24 panel review. 

We used the 1,115 proposals received in Cycle 23 as a training set to the 7 science categories used in Cycle 24: {\it Stellar Physics}, {\it Solar System}, {\it Extrasolar Planets and Planet Formation}, {\it Stellar Populations}, {\it Galaxies}, {\it Massive Black Holes and Their Hosts}, and {\it Intergalactic Medium and Cosmology}. These categories are slightly different from the subject areas used by the panels in Cycle 23 and earlier, and represent an effort by the SPG to reduce ambiguity in the sorting process. In Cycle 23 and earlier, the panels covered only 6 subject areas: {\it Planets}, {\it Stars}, {\it Stellar Populations}, {\it Galaxies}, {\it AGN and IGM}, and {\it Cosmology}. Also prior to Cycle 24, what were then called science categories were actually distinct from the subject areas of the panels which reviewed the proposals. In all there were 14 science categories in which proposers could apply to, which were divided into the 6 subject areas of the panels. 

To make an appropriate training set for the Cycle 24 testing, we manually re-categorized each of the Cycle 23 proposals to the Cycle 24 categories. For the most part the re-categorizations were easy, as neither the name or the subject area had changed greatly. However, what was previously {\it Planets} had split into a panel on the solar system, and one on extrasolar planets.  Proposals on the intergalactic medium were no longer paired with proposals on active galaxies, rather paired with cosmology, and a new panel was formed covering massive black holes and their hosts, taking some proposals which might otherwise been sorted into galaxies. Each of these switches required special attention on the scientific categories chosen, and in some cases the keywords selected, and in more than a few cases, a thorough re-reading of the proposal, to determine which of the new categories was most appropriate.

Using this training on the Cycle 23 proposals, PACMan then categorized the 1,094 proposals received in Cycle 24, and tested those results against the SPG's manual sorting process. We also tested the suitability of the panelists to the same subject areas, using the Cycle 23 training on proposals on abstract bibliographies. The results of classifications for Cycle 24 proposals, as well as panelists for the Cycle 24 TACs, are detailed below.

 \subsection{Results by Category}

 \begin{figure}[t] 
   \centering
   \includegraphics[width=3.5in]{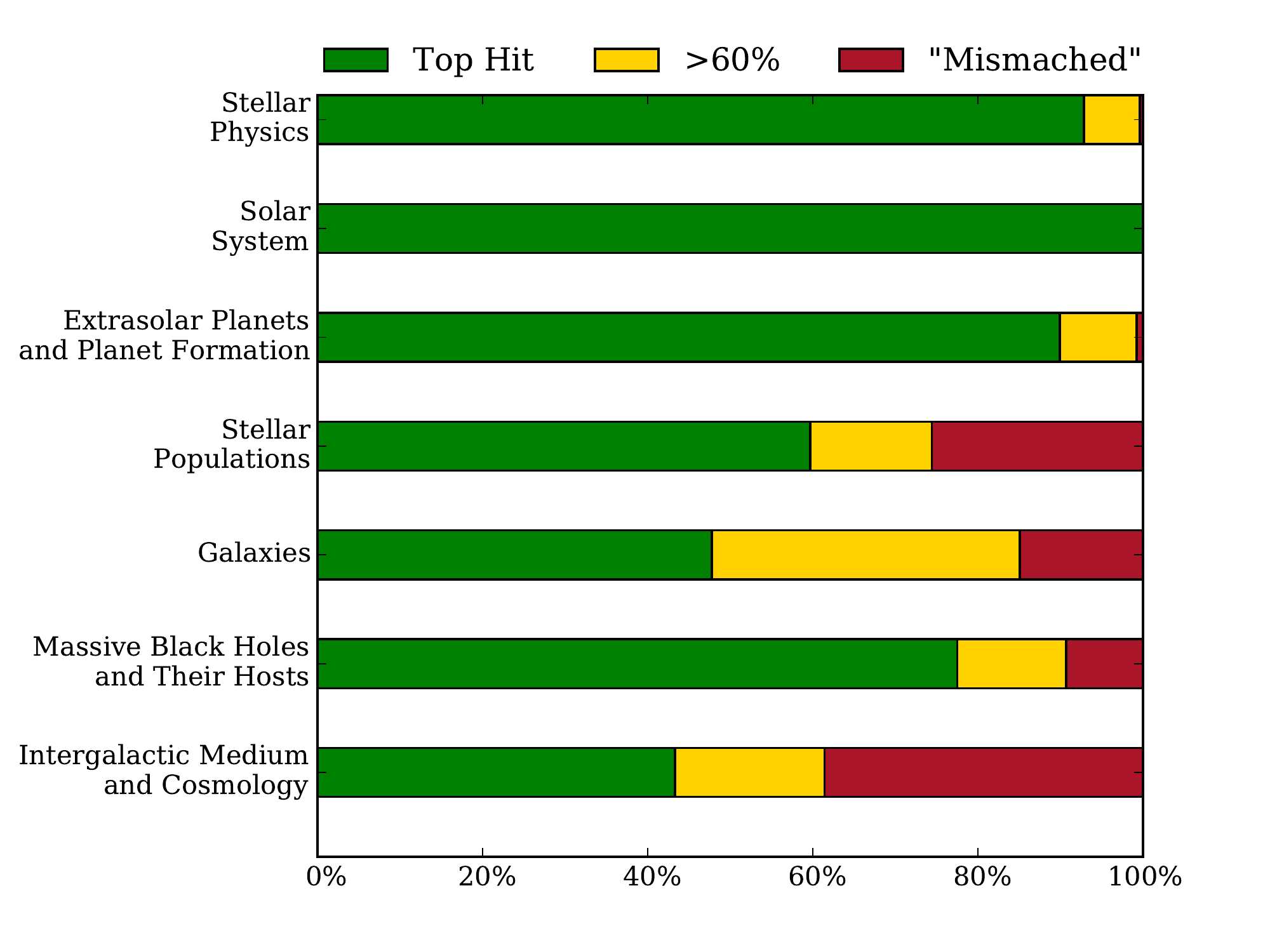}
   \caption{\footnotesize Comparison of PACMan categorization to the manual sorting process for HST Cycle 24 proposals. The percentage of proposals that are similarly categorized by both are given in green. Yellow indicates proposals whose manual classification is within the top 60\% of sorted PACMan classifications. Red proposals are categorized differently by each.}
   \label{fig:cytwentyfour}
\end{figure}

 A detailed breakdown of results for 7-category classification on Cycle 24 proposals is shown in Figure~\ref{fig:cytwentyfour}.  In the figure, the green bars show the percentage of proposals were given the same categorization, by PACMan's top hit and by proposers and the SPG (see more on this below). The yellow bars show the fraction of proposals for which the manual classification was within the 60\% most likely PACMan categorizations--- a marginal success for PACMan. And the red fractions indicate \textit{mismatches}, where the SPG-assigned categories fell into the bottom 40\% of PACMan's categorizations.

Perhaps not surprisingly, PACMan showed the highest success rates (i.e., showed most congruence with manual sorting) in categories with word-tokens that unambiguously belong to only said categories. For example, `Kupier' and `Venus' only appear in {\it Solar System} proposals, `T Tauri' in {\it Stellar Physics}, and `Coronagraphy' typically in {\it Extrasolar Planets and Planet Formation}. In contrast, the least successful categories, {\it Stellar Populations}, {\it Galaxies}, {\it Intergalactic Medium and Cosmology}, have long been areas of great ambiguity to proposers. Topics on star-formation histories, for example, have considerable overlap between {\it Stellar Populations} and {\it Galaxies}, and topics on galaxy formation and evolution have relevance to both {\it Galaxies} and {\it Intergalactic Medium and Cosmology}.

The SPG has always resolved potential misclassifications by reviewing the categorizations of all proposals manually, typically finding about 10\% should be moved (or swapped) to another category. Most of these swaps occur in the categories of stellar populations, galaxies, and cosmology, which show a lot of overlap, and sometimes are proposals with multiple, wide-ranging science goals. It is for this reason we do not assume inconsistent PACMan results are necessarily incorrect assignments. 

 \begin{figure}[t]
   \centering
   \includegraphics[width=3.4in]{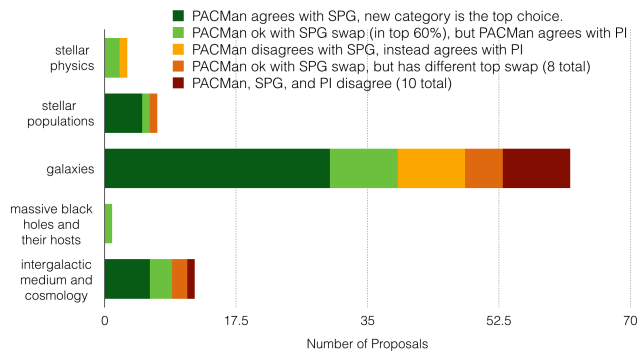}
   \caption{\footnotesize Comparing the resorting (or category swaps) of Cycle 24 proposals made by the SPG, to PACMan categorization results. 48\% of SPG swaps would have been predicted by PACMan, while another 28\% were in PACMan's 60\% most likely categories. On the remaining 24\%, PACMan disagreed with the SPG swaps, about half of which PACMan would have agreed with the proposers (PI) original categorization.}
   \label{fig:spg_comp}
\end{figure}
We show this by reviewing the PACMan results for the proposals the SPG independently recategorized (or swapped) between categories in Cycle 24. In Figure~\ref{fig:spg_comp} we show the number of proposals swapped by the SPG, indicating how many swaps were to PACMan's top choice categories (48\% of 85 swapped proposals), how may were to categories in PACMan's top 60\% (76\%), and how many PACMan did not agree with (24\%). It also shows the number of swapped proposals in each category that were most consistent with the proposers original classification (31\%), and which were assigned differently by PACMan, the SPG, and the proposers (12\%). A review of these last few proposals showed they indeed covered multiple goals, in broad areas, for which the primary science would have been somewhat ambiguous if not explicitly stated.

If used instead of SPG classification, PACMan would have ultimately made appropriate re-classifications, as indicated by Figure~\ref{fig:spg_comp}. Most disagreements between PACMan and SPG proposal swaps occurred either in borderline cases, where PACMan should be less subjective than the SPG, or in categories with weaker training sets and wider distribution, which can be remedied by using larger, hand-classified training sets.  With swaps, and accepting PACMan's top choice categories, we find this trained classifier adequately provided proposal sorting to the Cycle 24 categories 87.8\% of the time, from just the Cycle 23 training data. As will be shown in the next section, This is not as high as we might have expected from a $\sim1,000$ proposal training set ($>92\%$ success). The differences are likely attributable to the significant change in the panel science categories, which may have induced errors in appropriately sorting Cycle 23 proposals to Cycle 24 categories-- similar to what proposers themselves go through when proposal categories are changed.

 \subsection{Expected Improvements From Multiple Cycles of Proposals and Multiple Category Pools}
\begin{figure}[t] 
   \centering
   \includegraphics[width=3.5 in]{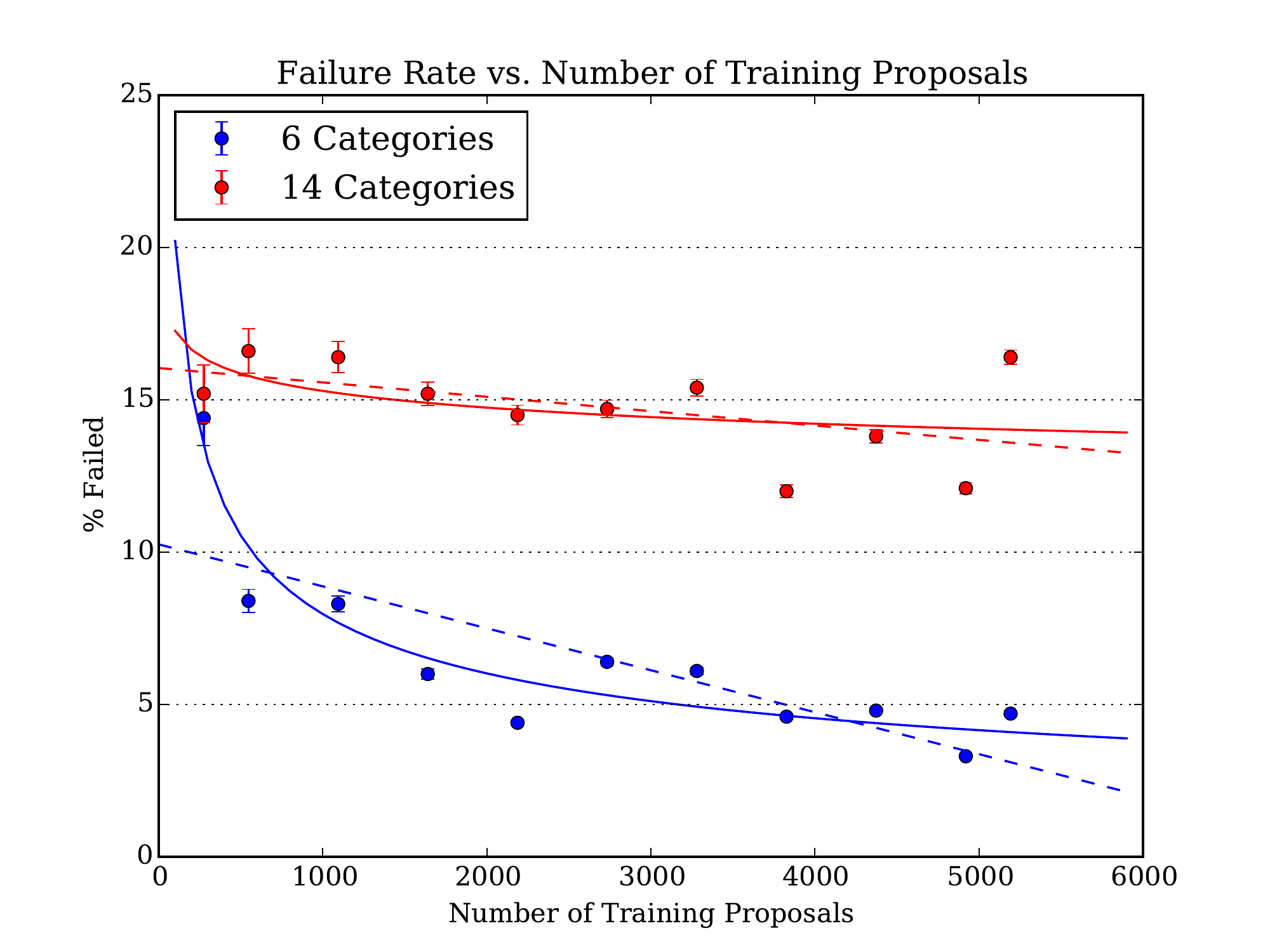}
   \caption{\footnotesize Misclassification (or failure) rate as a function of the number of training proposals for 6 and 14 category pools. Limiting classification to only 6 categories provides the best successes, which are likely achieved after training only $\sim2,000$ proposals.}
   \label{fig:cats_tests}
\end{figure}

This type of machine learning categorization should improve with larger training sets, and could broadly provide better categorization to more pools when numerous pools are used. We compiled the 5,461 proposals submitted over HST Cycles $19-23$, re-categorizing each to a common set of 6 science categories, which to our benefit had not changed significantly over that same 5 year period. We then ran a series of tests, each time using an increasing fraction of randomly selected proposals, from 5\% to 95\%, to create training dictionaries that are tested on the remaining sets of proposals, checking the success (or failure) rates of PACMan against human categorization. 

As can be seen in Figure~\ref{fig:cats_tests}, misclassification (or failure) rates were highest when the fewest training tokens were used-- 274 training proposals yielded an accuracy of about 86\%. The rate of accuracy rises with larger training sets, to around 96\% when  $\sim90\%$ of the available proposals are used for training. The uncertainties shown in Figure~\ref{fig:cats_tests} are Poisson errors, and do not reflect the potential arbitrary uncertainty from random selections which may favor one category pool over another, both in the training and in the test sets, which may invoke as high as 14\% error.

The trend in training improvement may represent a steady rise of about 1 percentage point per 725 training proposals or so, exemplified by the dashed blue line in Figure~\ref{fig:cats_tests}, suggesting that significant enhancements in the tool can be met with each new cycle of training data. However, the trend in the figure is more likely to be something like a power law, exemplified by the solid blue line, where most of the advantage in classification is met in the first $\sim2,000$ training proposals, and an inherent success floor is met, around 95 to 96\% for 6 categories, with little-to-no advantage seen in increasing the training set further.  

This model is seemingly descriptive of what is seen when the number of category pools is increased from the 6 that described the panel topic categories, to the 14 which were the scientific categories made available to proposers in those cycles. The red points and lines show the analogous results on 14 pools, exhibiting a much shallower improvement in failure rates with the number of proposals in the training set. It would seem that for the 14 science categories, sorting to an accuracy greater than approximately 85\% would be increasingly elusive with our method.

\subsection{Results on Categorizing Reviewers}

As detailed in Section~\ref{sec:reviewers}, PACMan can be used to categorize potential panelists by evaluating corpora constructed from ADS abstracts, in much the same way PACMan evaluates proposals, with one important distinction. PACMan categorization probabilities are treated as raw scores for authors, and not normalized as with the proposals. This allows for more of an evaluation of appropriateness to each category rather than forcing it into a discrete pool. However, to address this metric of appropriateness, it became important to re-evaluate the range in validity of the PACMan score.

\begin{figure}[t] 
   \centering
   \includegraphics[width=3.5in]{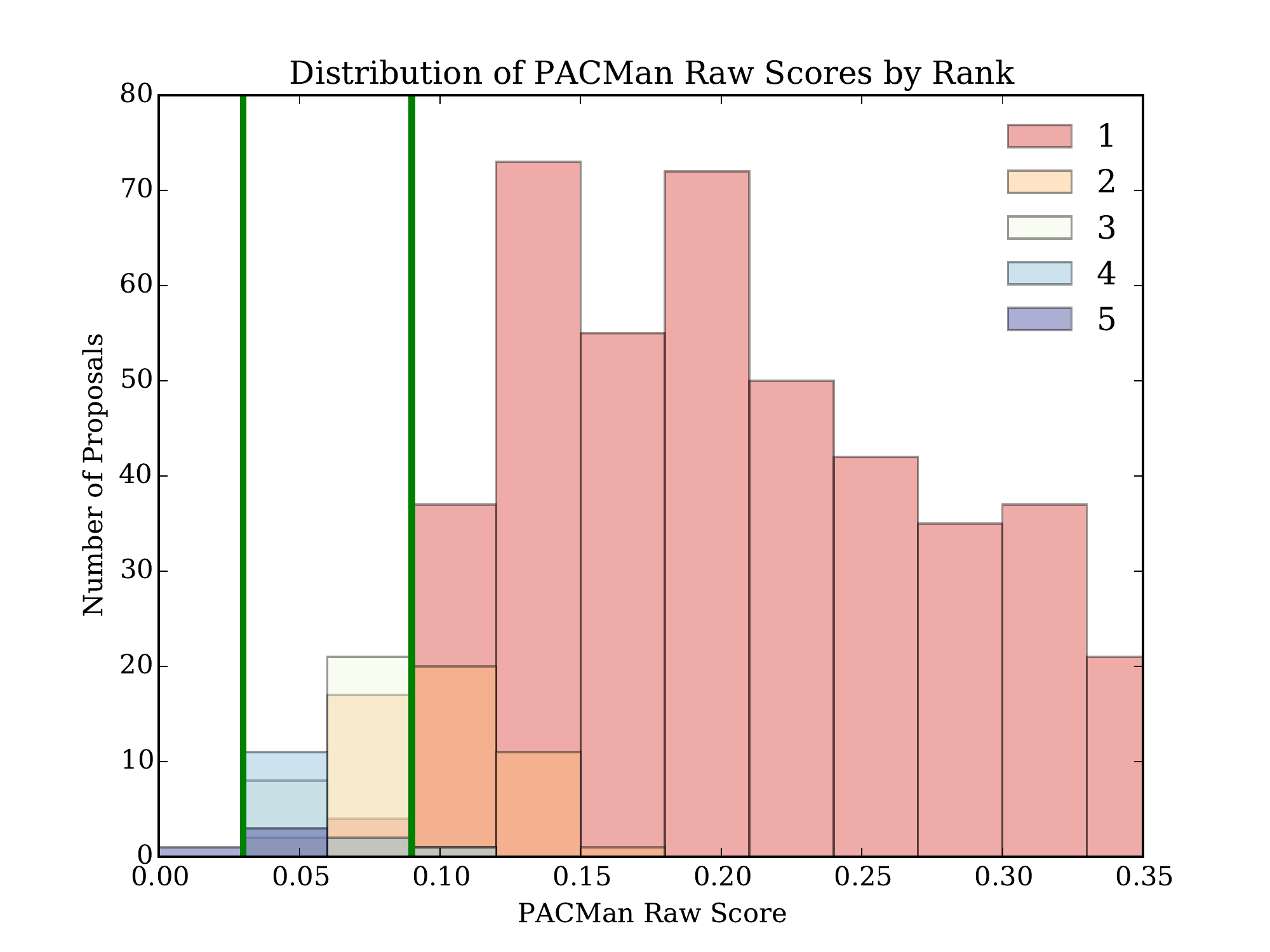}
   \caption{\footnotesize Histograms of PACMan raw scores (unnormalized) on Cycle 24 proposals. Scores are sorted by rank such that those with manual classifications in agreement with the $1^{st}$ ranked score are 1's (in red), those matched to the 2$^{nd}$ ranked score are 2's, and so on until 5's are matches in the last 3 scores. The figure shows the range in validity in PACMan raw scores, where $\ge0.03$ are probable categorizations,  $\ge0.09$ are highly likely, and $<0.03$ are unlikely.}
   \label{fig:hkap_scores}
\end{figure}

Despite the potential for wide granularity in the Robinson combination method, PACMan scores were empirically confined to about half the range of values, from 0.0 to as high as 0.6, with most in the range $0.1-0.3$. The histograms in Figure~\ref{fig:hkap_scores} show this restricted range in scores for the proposals tested in the Cycle 24 analysis, trained from Cycle 23 proposals. These are numerically represented and color-coded by rank such that a `1' indicates proposal whose first ranked categorization matched the manually assigned category, `2' are those whose second ranked category matches the assignment, and so on to `5' which were for matches to the last three ranked segments. 

It is interesting to note the range in scores for successfully sorted proposals is relatively large, taking most of the dynamic range in all scores from roughly 0.06 to 0.6. By contrast, the least successfully categorized proposals take up a very narrow range in scoring, from 0 to 0.06. Marginal success is similarly confined to a small range, between 0.03 to 0.15. There is no indication yet of how this score is impacted by the number of words within a corpus, while there is some expectation that they should be in some way correlated. Our corpora have very similar lengths, largely due to strict limits to the number of pages a proposal is allowed to devote to the Scientific Justification section. We leave this as a caveat to this study, and a test for future investigations.

We arbitrarily set two dividing lines, at 0.09 and at 0.03, indicating boundaries where `1' and `2' proposals make up more than half of scores, and '5' make up more than half of scores, respectively. PACMan raw scores on category above 0.09 are therefore interpreted as highly likely categorizations, and those below 0.03 are considered unreliable (or poor) categorizations.

When applied to the abstract bibliography of potential panelists, the scores are interpreted as a level of expertise in subjects covered by the category. Panelists whose scores for a given category are above 0.09 could be considered experts in the category, where as those whose scores are lower than 0.03 for a given category would be considered a less than ideal reviewers in that area.

\begin{figure}[t] 
   \centering
   \includegraphics[width=3.5in]{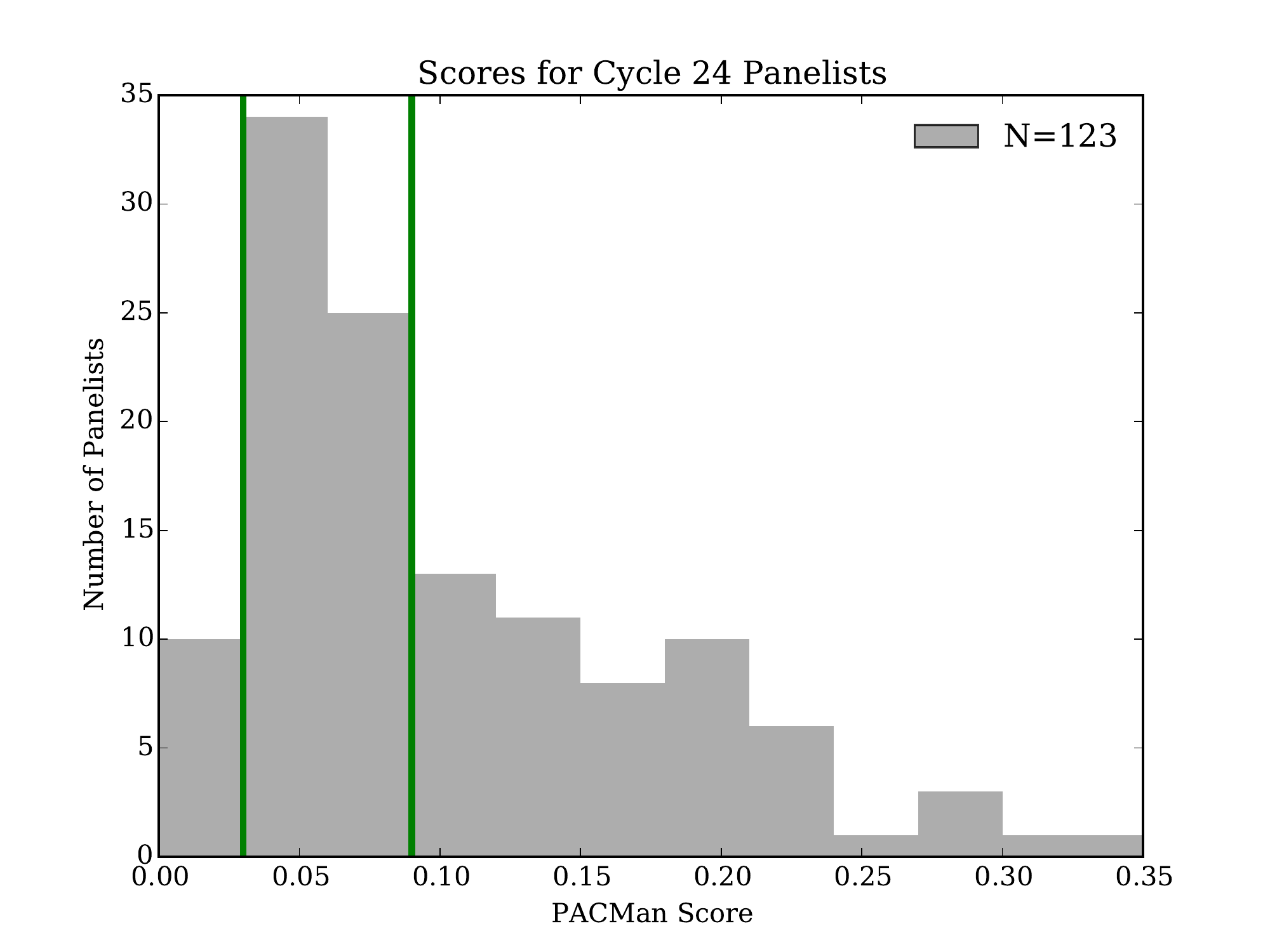}
   \caption{\footnotesize Histogram of PACMan raw scores for reviewers on the HST Cycle 24 TAC panels. Range of validity markers are the same as used in Figure~\ref{fig:hkap_scores}.}
   \label{fig:panel_scores}
\end{figure}

We performed a check of the PACMan scores for the 123 Cycle 24 TAC panelists (excluding the panel chairs, and at large members) in the categories for which they served--- the results of which are seen in Figure~\ref{fig:panel_scores}. As shown, 44\% of panelist who served were considered experts in their panel category by the software. 92\% were considered to have moderate expertise or better in their area, and only 8\% (10 panelists) were considered less than ideal reviewers for their review area. Although small, it would seem a bit surprising this is not zero, but there are two good explanations for this. First, as discussed in previous sections, there is a minimum number of word tokens a corpus needs to be successfully categorized. Tests have shown this floor is approximately 3,000 words. Panelists with few publications in their abstract bibliography (but perhaps with significant impact--- hence their recruitment) will generally receive misleading PACMan scores, tending towards poor ranks.

Secondly, recruiting for the TAC panels service is one of the most arduous tasks the SPG does in organizing the review. Identifying available and willing panelists for this large review year after year requires significant effort, especially when one considers potential conflicts of interest candidates may have in these increasingly collaborative fields. The SPG is also subject to the same conflict of interest issues as panelists, and do not participate in the construction of panels in their primary science areas. As such, identifying reviewers for subjects outside our areas of expertise can sometimes result in less than ideal recruits. This, however, has not been detrimental, as experience has shown that these reviewers have provided just as rigorous reviews as their more experienced counterparts, and also often bring `relevance to the bigger picture' critiques that would otherwise be missing in these discussions. It also bears mentioning that proposers are cautioned that proposals are evaluated by panelists who are not necessarily experts in their field, and that scientific justifications should be written with this in mind. It is also important to note that proposals receive grades from multiple reviewers, both at preliminary and final stages, and so the impact of non-experts on the final performance of a proposal is somewhat diluted. 

Our goal for this automated tool is to more objectively apply it to a large list of investigators, not just those listed in ProPer, affording us a wider pool of potential reviewers with relevant expertise in the science, regardless of experience with HST.

\section{Discussion and Future Directions}
As we have shown, PACMan is a capable tool for replicating key tasks in creating peer reviews for time with HST. Our testing based from a single cycle of training data showed effective sorts into the existing science categories, making the same sorting decisions for the Cycle 24 proposals as the SPG an average of 87\% of the time. We found that nearly all of the panelists (92\%) selected to serve in the Cycle 24 review were deemed appropriate reviewers for the subject area by PACMan, nearly half of which (44\%) were considered experts in their area of review.

Tests over 5 prior cycles of proposals, in a period when science categories were more similar to one another, shows improving success rates with larger corpora of test data, approaching an average of 95\% to 96\% after only a few cycles of training data, for six (and likely seven) science categories. It is reasonable to expect this level of accuracy with PACMan for Cycle 25 and future cycles with HST, and perhaps with JWST, but this will require non-trivial effort to re-categorize past proposals over several cycles to the current scientific category definitions for the dictionary training. Moreover, each time the category definitions change in the future, either in response to perceived changes in disciplines or to somehow improve the organization of the process, the training set will have to be re-categorized to match the new definitions. This will be important to consider when assessing the net effort saved using PACMan sorting in the first place.

PACMan showed a path to more efficient and inclusive selection of potential panelists. The ADS abstract bibliography searches, in most cases, provided useful corpora to test against the proposal-trained dictionaries, and test the appropriateness of a reviewer to a specific review subject. However, while not described in previous sections, there were indeed frequent issues in extracting a unique and complete bibliography due to limitations in ADS name searches. This type of extraction would be much improved by using more persistent and distinguishable identifiers, like ORCID~\footnote{see {\tt http://orcid.org}.} identifiers, rising in popularity. And while the abstracts alone may be sufficient, it would be more ideal to use the text of each published manuscript, which at this time is not as readily available in electronic searches as the abstracts. Lastly, with larger and more complete corpora, the results can be properly weighted by higher-level information on the quality of each publication, e.g., rank in authorship list and citation history, be further refine the interpretation of appropriateness to a given science category.

PACMan reinforced a known issue with proposals that do not uniquely favor one categorization over another. These cross-category, often multi-disciplinary, proposals may hint at the need for some reorganization of the science categories, better suiting these misfit  proposals, and perhaps also providing a better load balance across panels.  It is not often obvious before hand if a reorganization is necessary, and what the new categories should be. But an analytical tool may be able to predict such changes in advance. One can conceive of a mechanism for creating `fluid pools'--- an automated self-organization of corpora based on their subject matter, perhaps involving Markov-Chain Monte Carlo (MCMC) applications of PACMan to find the most advantageous sorting to an undefined number of pools. One could imagine training the MCMC as we have done for PACMan to the defined categories, over several cycles of data to not only find the best pool definitions, but the appropriate training to those new pools. Reviewers would similarly be selected for expertise in the new pools.

Throughout this endeavor we have sought to automate a proposal and proposer classification routine to about the $90\%$ accuracy level, which PACMan has done successfully. However, we acknowledge that a simple naive bayesian approach is not state-of-the-art in realm of corpus classification with machine learning. There are several modifications one could make, beyond the ones we employ in PACMan, as well as many other methods that exist outside the scope of this project which may, when implemented in future iterations of PACMan, further improve our results. 

Among these modifications is the Multinomial Naive Bayes method, which more stringently relies on the frequency of a word's usage in training sets to arrive at the classification, assuming the probability of a corpus belonging to a specific class, based on the given word, is the multinomial distribution~\citep{MN1998}. This would effectively replace Equation~\ref{eqn:pj} with:

\begin{equation}
	P_j=\prod_i^N\frac{P(w_i|p_j)^{V_{i}}}{V_{i}!},
	\label{eqn:pj2}
\end{equation}
\noindent where $V_{i}$ is the number of times word $w_i$ appears in the corpus. Given that the multinomial approach also provides the granularity that a simple naive probability lacks, it likely eliminates the need for the Robinson modification we have used in PACMan, eliminating Equation~\ref{eqn:pj1}. And as the distribution of word use is more important than its raw frequency, this method may be better in dealing with small corpora (like abstracts) with relatively few word-tokens.

The Term Frequency-Inverse Document Frequency (TF*IDF) is yet another potentially simple modification to provide additional weights in determining a word-token's importance in classifying a corpus. The method combines two statistics, the Term Frequency which applies some weight to words used frequently in corpora of a given class, and Inverse Document Frequency, which essentially down-weights words that commonly appear in all corpora~\citep{SR2004,jones1972}. A potentially easy application of TF*IDF in PACMan could be to replace the trigger word, or $P(w_i|p_j)$ in Equation~\ref{eqn:pij}, with the product of a TF$(w_i)$, from the frequency  of use of the word in training corpora, and an IDF of the form
\begin{equation}
	{\rm IDF}(w_i)=\log(N_{cat}/n_i),
\end{equation}
\noindent where $n_i$ is the number of corpora the word-token appears in, and $N_{cat}$ is the number of corpora in the training catalog. This type of TF*IDF would also supplant the use of stop-words in our routine. 
 
A different direction entirely from Bayesian methods, Density-based spatial clustering of applications with noise, or DBScan, would parse corpora and cluster them based on their similarity to a given category in the training set~\citep{E1996}. Such a routine would then draw classification ``boundaries'' around the densest points of these clusters. Cluster boundaries could then be re-evaluated and adjusted for each category-- somewhat akin to what is done in neural networks.

But focusing more on the outcome than the method, the holy grail in this type of work would be a mechanism to more directly match proposals to a reviewer or select pool of reviewers. We have tested a method of training a dictionary on a given proposal, against other proposals in the same category, then testing that trained dictionary on a potential reviewer's abstract bibliography, to test relative appropriateness for reviewing the test proposal. The results of these tests were encouraging but problematic--- the method would routinely select the same, most prolific authors in a given science area as the most ideal reviewers. This probably not surprising, given PACMan's current reliance on non-informative priors.  Here, it may be useful to include priors on the quality of the publications (e.g., rank in authorship and citations) for improved matches. One could also imagine using other priors such as the number of other assigned proposals, or the frequency of previous service in STScI committees in general, as ways to balance the load and to work towards greater impartiality.

\acknowledgements Thanks to Brett Blacker, Andrew Fruchter, Janice Lee, Claus Leitherer, Jenifer Lotz, and Amaya Moro-Mart{\'i}n for their discussion in preparing and evaluating PACMan, and to the STScI Space Astronomy Summer Program for funding much of this effort.

\end{document}